\documentclass[times,twocolumn,final,authoryear]{elsarticle}

\usepackage{jasr}
\usepackage{framed,multirow}

\usepackage{amssymb}
\usepackage{latexsym}
\usepackage{caption}
\usepackage{subcaption}

\usepackage[switch]{lineno}

\usepackage{url}
\usepackage{xcolor}
\definecolor{newcolor}{rgb}{.8,.349,.1}

\usepackage[citebordercolor=white]{hyperref}
\journal{Advances in Space Research}

\begin{document}

\verso{Given-name Surname \textit{etal}}

\begin{frontmatter}

\title{Atmospheric waves disturbances from the solar terminator according to the VLF radio stations data}

\author[1]{Oleg \snm{Cheremnykh}}
\ead{oleg.cheremnykh@gmail.com}
\author[1]{Alla \snm{Fedorenko}}
\ead{fedorenkoak@gmail.com}
\author[1]{Anna \snm{Voitsekhovska}}
\ead{voitsekhovska.anna@gmail.com}
\author[1]{Yuriy \snm{Selivanov}}
\ead{yuraslv@gmail.com}
\author[2]{Istvan \snm{Ballai}}
\ead{i.ballai@sheffield.ac.uk}
\author[2]{Gary \snm{Verth}}
\ead{g.verth@sheffield.ac.uk}
\author[3]{Viktor \snm{Fedun}\corref{cor1}}
\cortext[cor1]{Corresponding author:}
\ead{v.fedun@sheffield.ac.uk}

\address[1]{Space Research Institute, 40 Acad. Glushkov Ave., Kyiv 03187, Ukraine}
\address[2]{Plasma Dynamics Group, School of Mathematics and Statistics, The University of Sheffield, Hicks Building, Hounsfield Road, Sheffield, S3 7RH, UK}
\address[3]{Plasma Dynamics Group, Department of Automatic Control and Systems Engineering, The University of Sheffield, Mappin Street, Sheffield, S1 3JD, UK}

\received{XXX}
\finalform{XXX}
\accepted{XXX}
\availableonline{XXX}
\communicated{XXXX}

\begin{abstract}
The perturbations from the solar terminator in the range of acoustic-gravity waves (AGWs) periods from 5 minutes to 1 hour were analysed with the use of  measurements of VLF radio signals amplitudes on the European radio path GQD--A118 (Great Britain--France). These observations provide information on the propagation of waves at altitudes near the mesopause ($\sim$ 90 km), where VLF radio waves are reflected. On the considered radio path a systematic increase in fluctuations in the amplitudes of radio waves was observed within a few hours after the passage of the evening terminator. For April, June, October 2020 and February 2021 events, the distribution of the number of wave perturbations with large amplitudes over AGWs time periods has been studied. Our results show that the evening terminator for different seasons is dominated by waves in the range of periods of 15--20 minutes. The amplitudes of the AGWs from the terminator at the heights of the mesosphere (fluctuations in the concentration of neutral particles, velocity components and vertical displacement of the volume element) are approximately determined by the fluctuations of the amplitudes of the VLF radio signals. The amplitudes of the AGWs on the terminator are 12--14\% in relative concentration fluctuations, which correspond to the vertical displacement of the atmospheric gas volume of 1.1--1.3 km. Based on the analysis of the AGW energy balance equation, it was concluded that the waves predominantly propagate in a quasi-horizontal direction at the terminator. The possibility of studying the long-term changes in the mesosphere parameters using fluctuations in the amplitudes of VLF radio waves at the terminator is shown.
\end{abstract}


\end{frontmatter}


\section{Introduction}\label{s:1}

The solar terminator is a global source of various types of atmospheric disturbances \citep{Forbes2009}.  As indicated by theoretical studies \citep{Beer1973,Somsikov1983,Somsikov1995}, ground-based onservations \citep{Galushko1998,Afraimovich2009} and satellite observations in the Earth's atmosphere and ionosphere \citep{Forbes2008,Lizunov2009,Liu2009,Bespalova2016} it include acoustic-gravity waves (AGWs) as well. The possibility of generating atmospheric waves by the solar terminator for the first time was reported by \citet{Beer1973}.

The solar terminator can be described as the sharp boundary between the region of the atmosphere illuminated by the Sun and the Earth's shadow. The projection of the speed of the terminator onto the horizontal plane $V_{ST}$ is about 460 m s$^{-1}$ near the equator. This speed weakly depends on the height in the atmosphere and decreases in the direction from the equator to the pole with increasing latitude. The optical terminator, as the visible boundary between light and shadow, is not a direct source of AGWs. The so-called "physical" terminator, or the region of sharp gradients of atmospheric parameters, which arises as a result of the absorption of solar energy and moves approximately at the speed of the Earth's rotation, is considered to be a source of wave disturbances \citep{Somsikov1983}. The low latitudes near the geographic equator are most favourable for observing wave disturbances on the terminator, where they have the largest amplitudes and present for a long time \citep{Somsikov1983}. The numerical modeling of the generation of wave disturbances by a moving boundary between the solar illuminated area and the Earth shadow (terminator, solar eclipse) makes it possible to take into account the real features of the source \citep{Karpov2008, Kurdyaeva2021}.

Observations of disturbances associated with the solar terminator are obtained using ground-based observations of the ionosphere using various remote methods (ionospheric sounding, incoherent scattering, observation of the total electron content using GPS, etc., see e.g. \citet{Galushko1998,Afraimovich2009}). These observations allow us to follow changes in the ionised component but do not provide information about neutral atmospheric gas. On the other hand, disturbances on the terminator can also be observed using in situ satellite measurements. Low-orbiting satellites make 14--16 revolutions around the Earth per day, and, on each revolution, they cross the terminator $line$ twice in the morning and evening local time. However, such studies are rather limited due to the need to fulfill certain conditions regarding the height and configuration of the orbit, as well as scientific equipment \citep{Forbes2008,Liu2009,Bespalova2016}. AGWs on the terminator were previously studied based on measurements of the Atmospheric Explorer-E equatorial-orbit satellite in the interval of atmospheric heights of $250-400$ km \citep{Bespalova2016}.
In the present work, we investigated wave disturbances on the evening terminator using ground-based measurements of the amplitudes of radio waves of very low frequencies (VLF). Perturbations were studied in the range of periods corresponding to atmospheric AGWs from $\approx 5$ min (Brunt-V\"ais\"al\"a period) up to $\approx 30$ min. Propagation of VLF radio waves occurs in the Earth-ionosphere waveguide with a reflection height during the day at altitudes of $\approx 75$ km (D-region of the ionosphere) and at night at altitudes of $\approx 90$ km (E-region of the ionosphere), see e.g.  \citet{Yampolskij1984,Wait1964}. The global network of VLF receivers opened up wide opportunities for diagnosing the state of the lower ionosphere and mesosphere \citep{Silber2016}. VLF measurement data of radio stations can be used to solve a number of scientific problems, in particular, the study of the influence of sources of space and ground origin on the state of the lower ionosphere \citep{Silber2016}. The propagation of AGWs in the atmosphere is usually recorded in the form of periodic fluctuations in the amplitudes and phases of VLF radio waves. Such fluctuations with periods of tens of minutes can be clearly visible in nighttime measurements, as well as at the terminator \citep{Nina2013,Rozhnoi2014}.
\begin{figure}
\centering
\includegraphics[scale=.30]{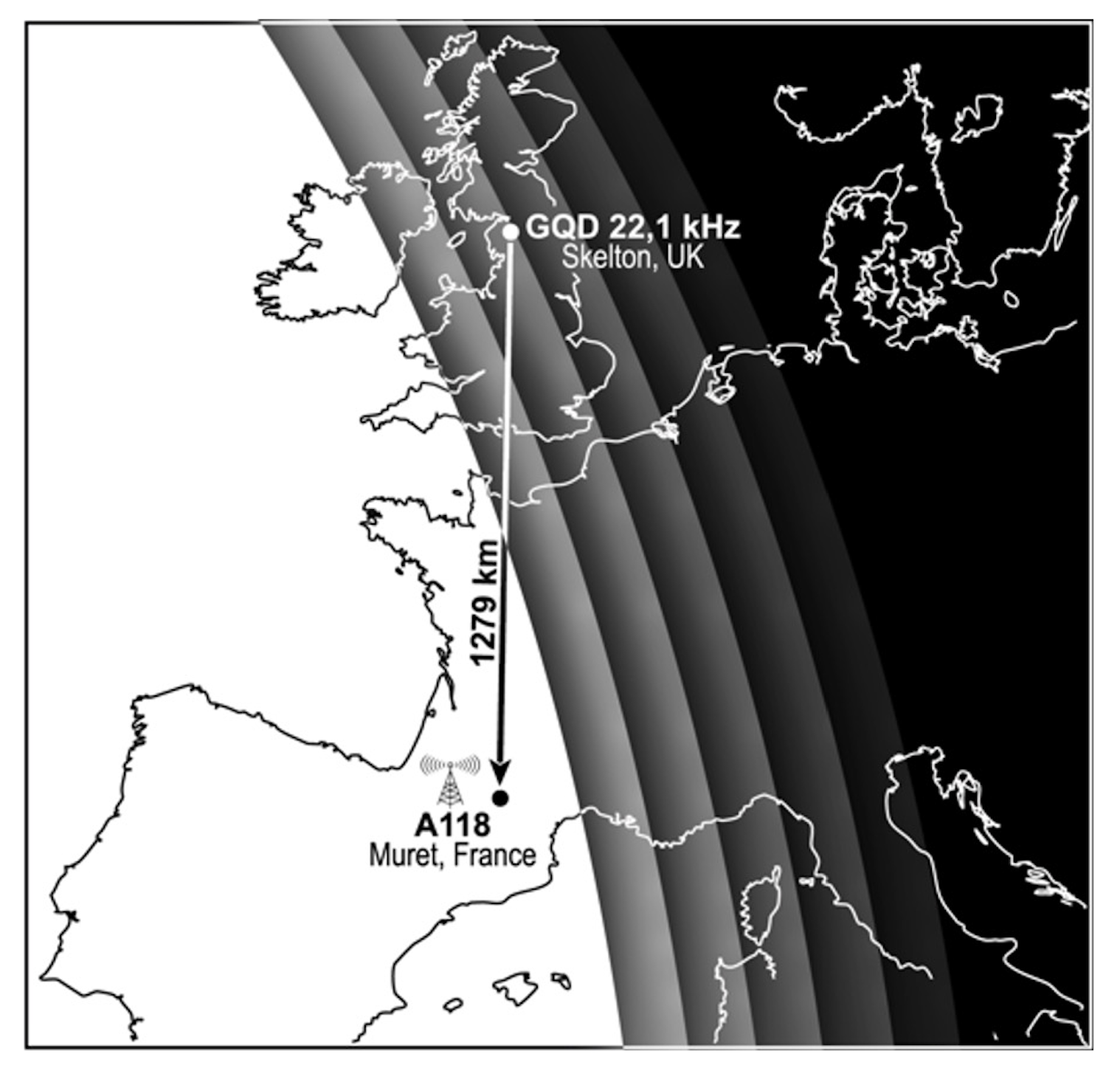}
\caption{The sketch of solar terminator line and wavefronts relative to the considered radio path GQD-A118.}
\label{FIG:1a}
\end{figure}

\section{Observational data analysis}\label{s:2}

\begin{figure*}
     \centering
     \begin{subfigure}[b]{0.45\textwidth}
         \centering
         \includegraphics[scale=.44]{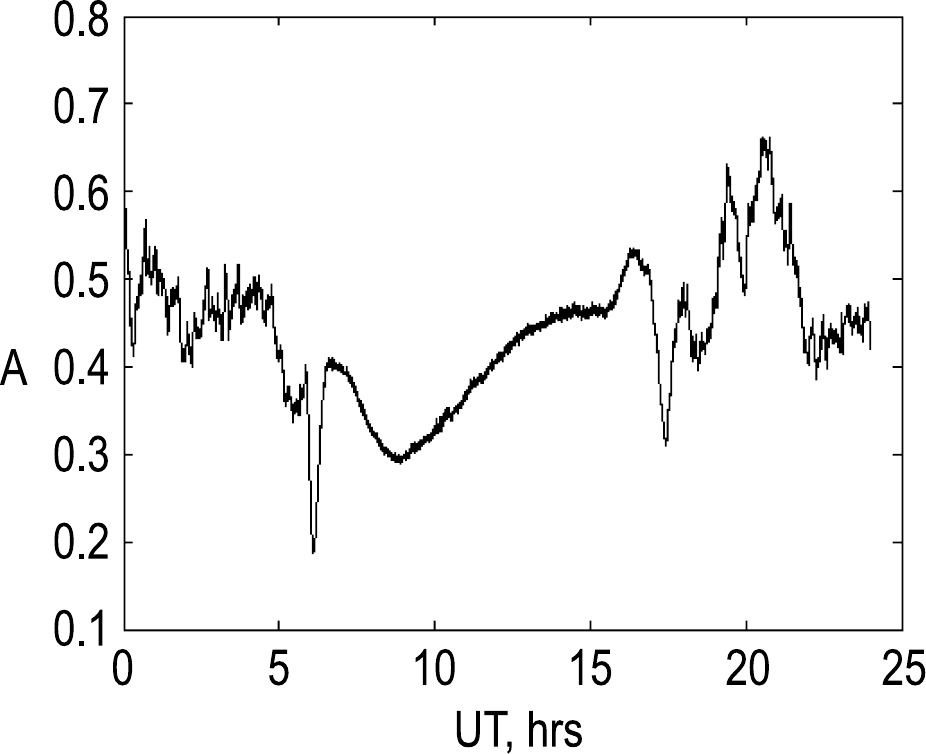}
         \caption{}
     \end{subfigure}
     \hfill
     \begin{subfigure}[b]{0.45\textwidth}
         \centering
         \includegraphics[scale=0.44]{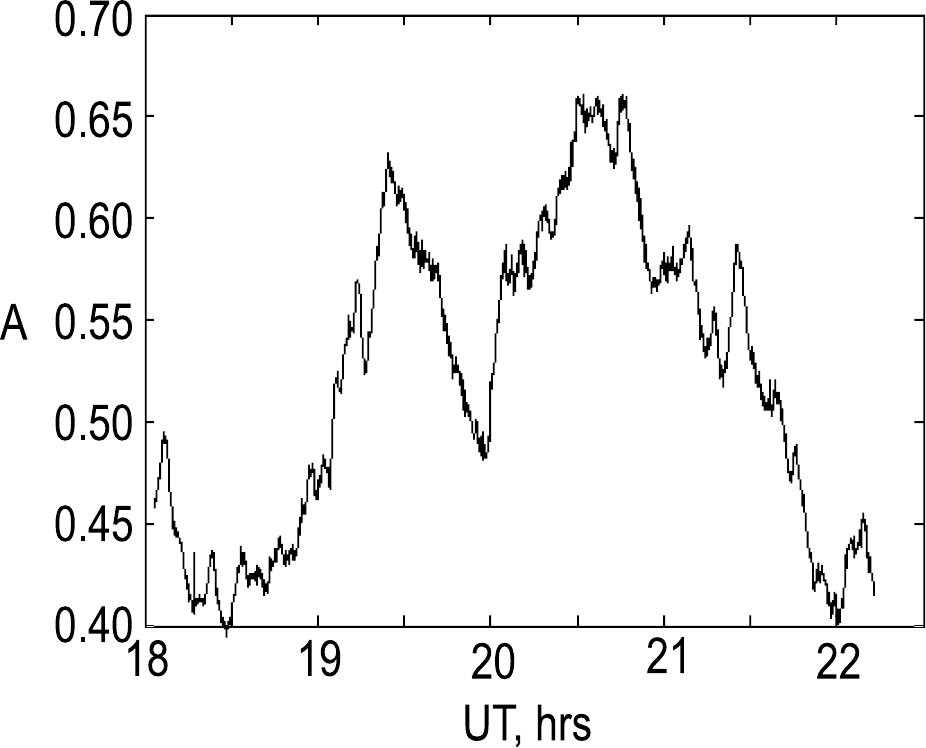}
         \caption{}
     \end{subfigure}
     \hfill
     \begin{subfigure}[b]{0.45\textwidth}
         \centering
         \includegraphics[scale=0.44]{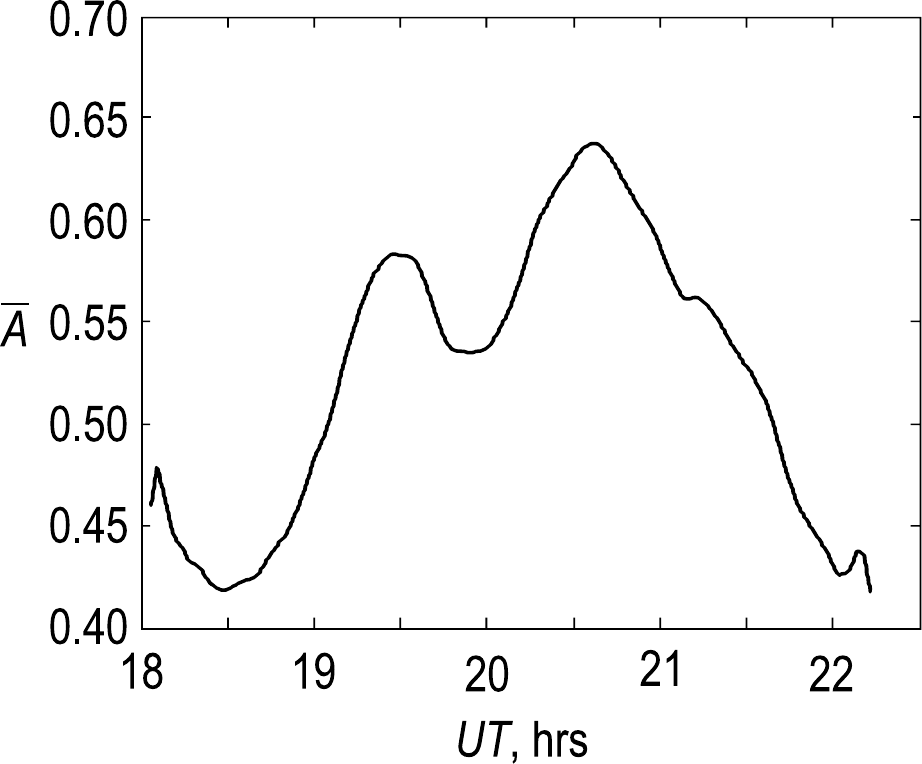}
         \caption{}
     \end{subfigure}
     \hfill
     \begin{subfigure}[b]{0.45\textwidth}
         \centering
         \includegraphics[scale=.44]{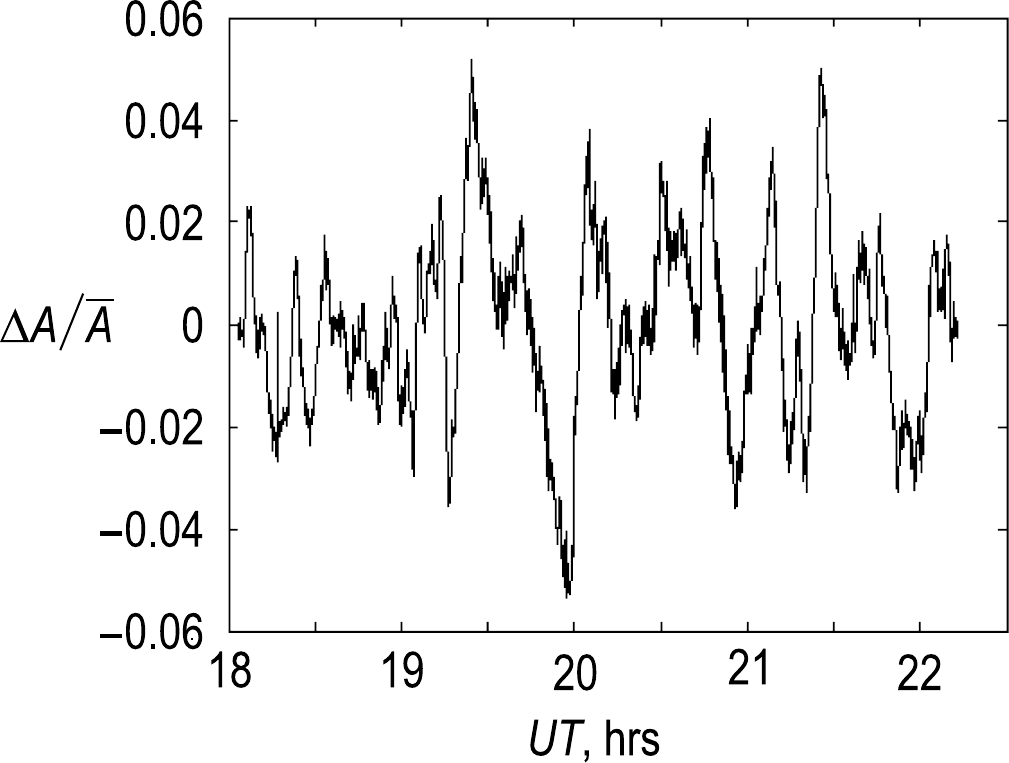}
         \caption{}
        \end{subfigure}
        \caption{Temporal variation of the amplitude of the VLF radio signal, $A$, on the GQD--A118 path on October 25, 2020 (a); The same variation of amplitude $A$ in the evening hours (b); the smoothed value of the amplitude, $\bar{A}$, by the moving average method (c); relative amplitude fluctuations, $\Delta A / \bar{A}$, in the evening hours (d).}
        \label{FIG:1}
\end{figure*}

To study the wave disturbances on the terminator in our work, we used the data from a VLF radio transmitter at a mid-latitude station in Great Britain (GQD, $f=22.1$ kHz) with a reception point in France (A118). The data sets (with the sampling rate $0.1$ s$^{-1}$) are available via https://sidstation.loudet.org/data-en.xhtml. The length of the considered GQD--A118 radio path is 1279 km. The location of the terminator and wavefronts relative to the selected radio path in the horizontal plane is shown in Fig. \ref{FIG:1a}. 
The use of geometric optics approximation is a suitable framework to study these waves at relatively short distances. i.e. less than 1500 km \citep[see e.g.][]{Yoshida2008,Fedorenko2021}.
Some properties of the AGWs can also be determined by measuring the amplitudes of radio signals. 
Due to the presence of sharp amplitude jumps associated with changes in the effective height of radio waves reflection, as well as conditions in the atmosphere after sunrise and sunset \citep{Yoshida2008,Fedorenko2021}, the data processing of VLF waves at the terminator is difficult. It is clearly seen in Fig. \ref{FIG:1} where the time dependence of the radio signal's amplitude along the GQD--A118 path for October 25, 2020 is shown. At the moment of passing the solar terminator, a sharp decrease in signal amplitude is systematically observed in the morning and evening. Note, that the nature of the change in the amplitude at the terminator differs on different tracks depending on the length of the track and the frequency of the radio signal, but these changes are always sharp \citep{Yoshida2008,Fedorenko2021}. Therefore, during the automatic processing of data series by using standard methods of spectral analysis, these sharp changes in amplitude provide strong non-physical spectral harmonics. Here were not interested in the moment of the passage of the solar terminator, but in the wave disturbances that accompany it and develop after the terminator passing. By assuming that the main transition processes from daytime to nighttime conditions have already occurred, the time interval was chosen to be far enough from the evening terminator passage. It is important for another reason as well, i.e. the theory of freely propagating AGWs (used for the numerical estimates later in the paper) is applicable for the modeling of the processes which are far away from the source of disturbances. In this regard, in order to search for waves from the terminator, we considered evening sections of data after the passage of the terminator lasting several hours (Fig. \ref{FIG:1}, panel (b)). At the same time, the very moment of passing the terminator was excluded from the analysis by assuming that the wave activity develops after it.

For the analysis of wave disturbances, it is necessary to separate them from large-scale changes of another origin. To achieve this, the output signal was considered in the form $A=\bar{A}+\Delta A$, where $\bar{A}$ is the average undisturbed value of the amplitude, and $\Delta A$ is the disturbance. To obtain a smoothed curve for $\bar{A}$, we applied the moving average method with a rectangular one-hour averaging window. The sizes of the averaging window were selected to single out the disturbances with periods less than 1 h in the original series. These disturbances correspond to medium-scale AGWs in the atmosphere.

The curve obtained by this method is shown in Fig. \ref{FIG:1}, panel (c). For wave processes, it is appropriate to consider relative fluctuations $\Delta A / \bar{A}$. This consideration excludes information on amplitude changes which are not related to the wave propagation, both of a physical nature and of technical origin, caused, among other things, by the peculiarities of the radio route and signal reception \citep{Fedorenko2021}. The obtained values of $\Delta A / \bar{A}$ for the evening hours of October 25, 2020 are shown in Fig. \ref{FIG:1} panel (d).

\begin{figure*}
	\centering
		\includegraphics[scale=.40]{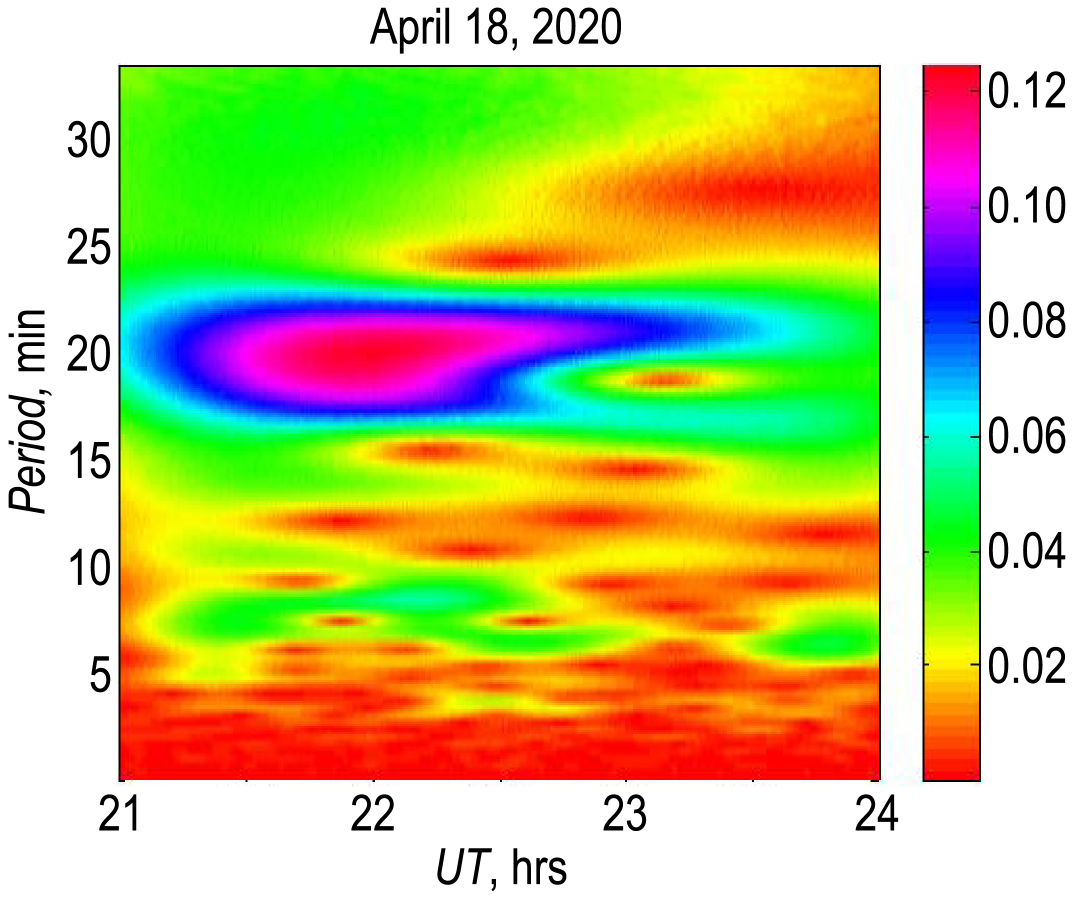}
		\includegraphics[scale=.40]{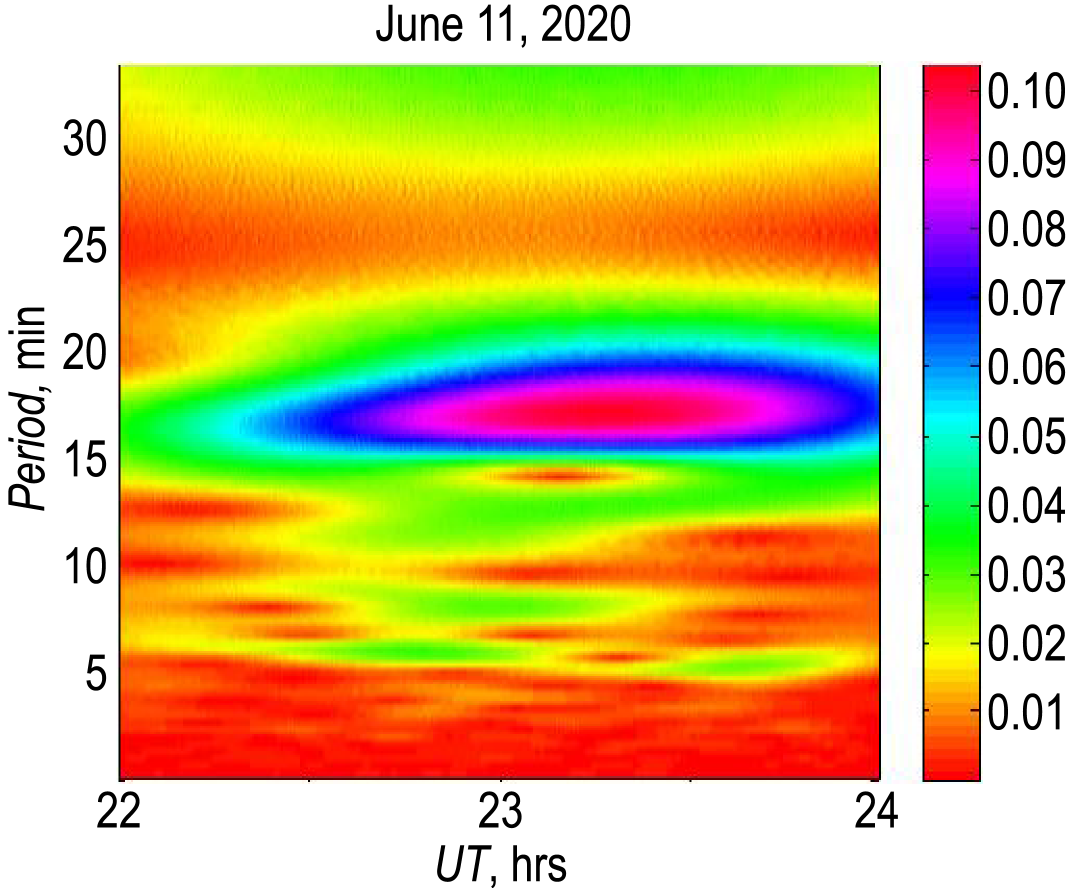}
		\includegraphics[scale=.40]{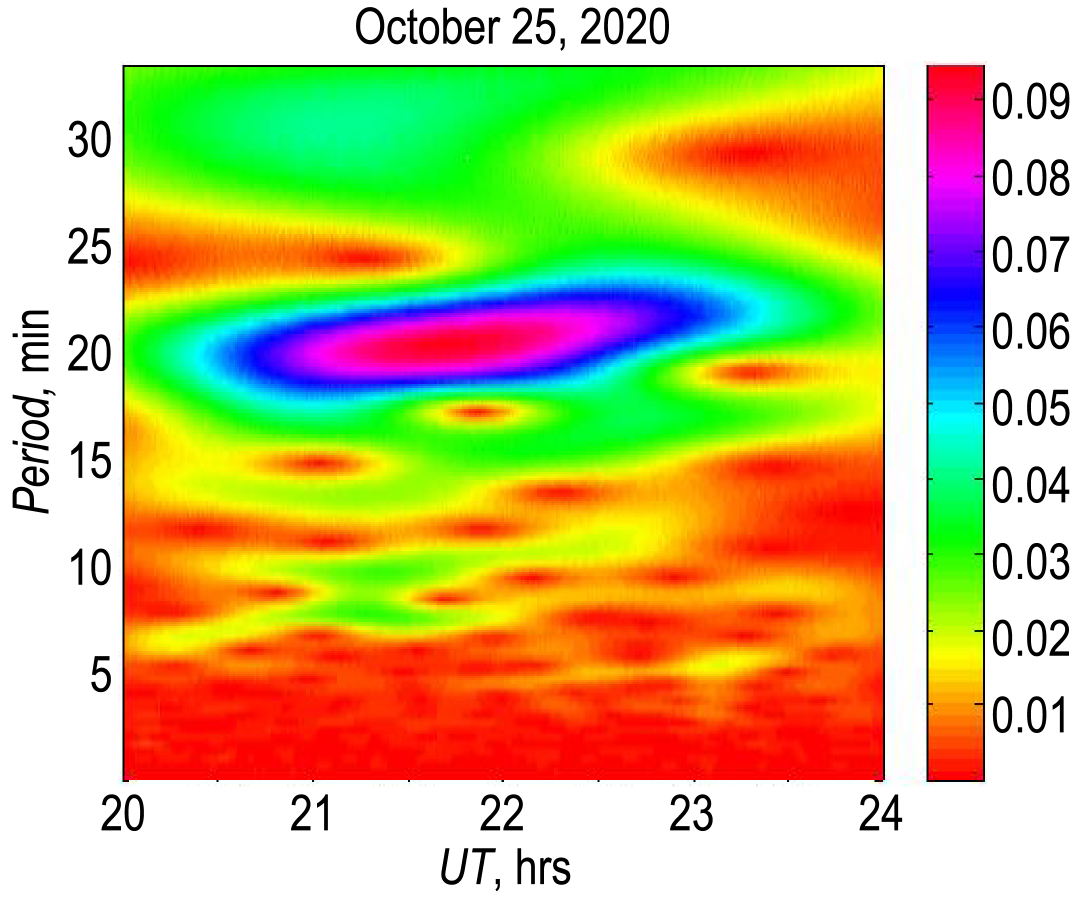}
		\includegraphics[scale=.40]{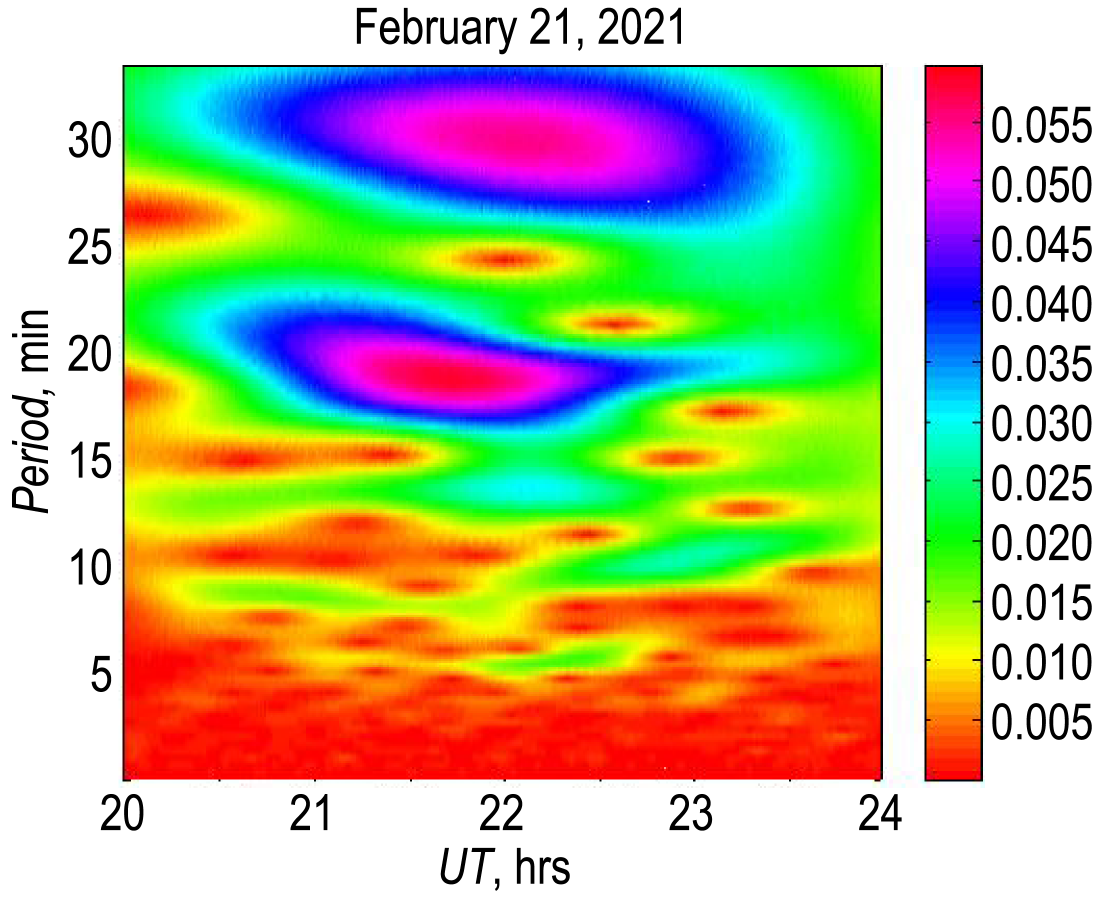}
	\caption{Amplitude wavelet spectra of waves disturbances at the evening terminator along the GQD--A118 path on separate days corresponding to different seasons.}
	\label{FIG:2}
\end{figure*}

\begin{figure*}
	\centering
		\includegraphics[scale=.41]{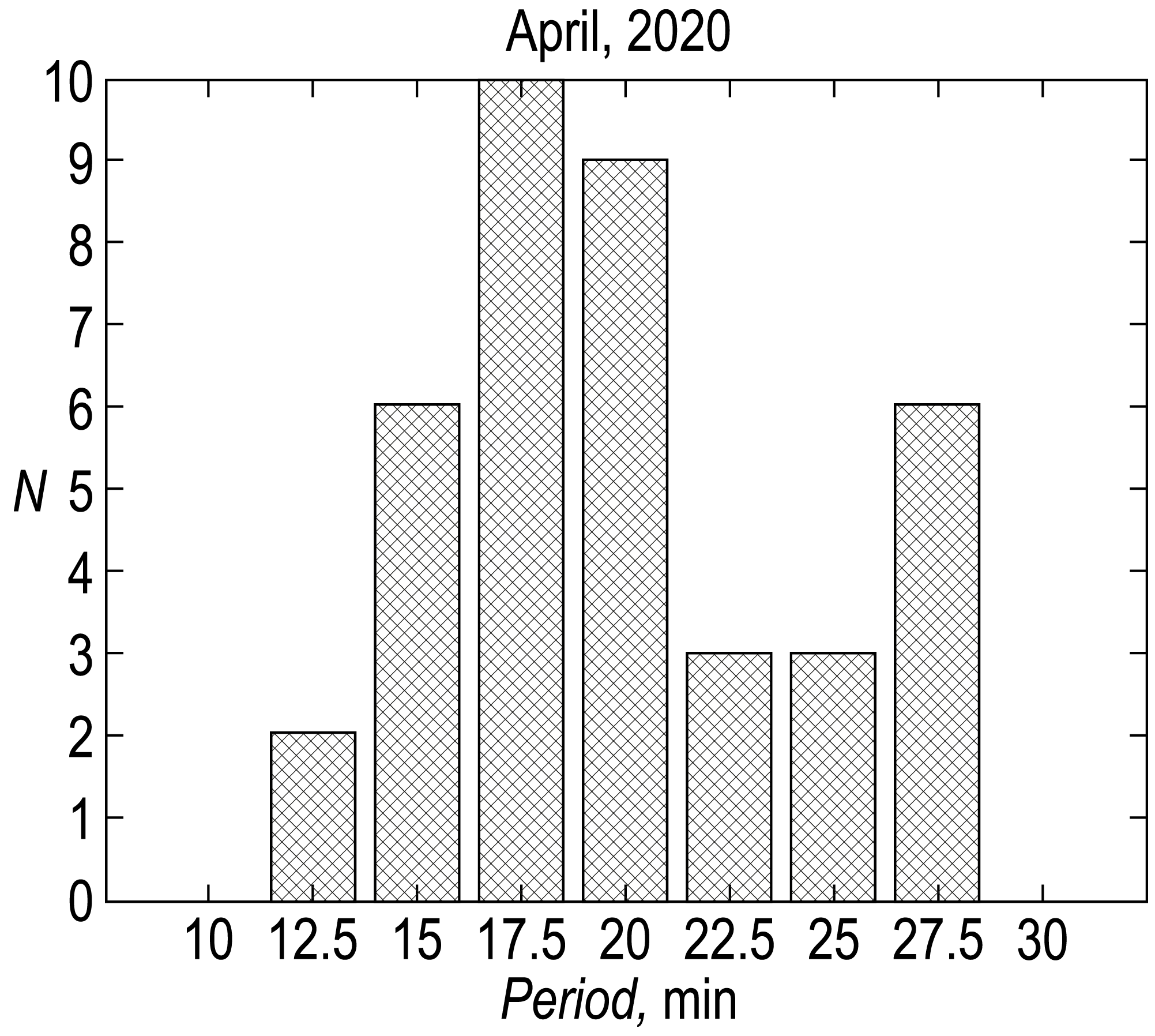}
		\includegraphics[scale=.41]{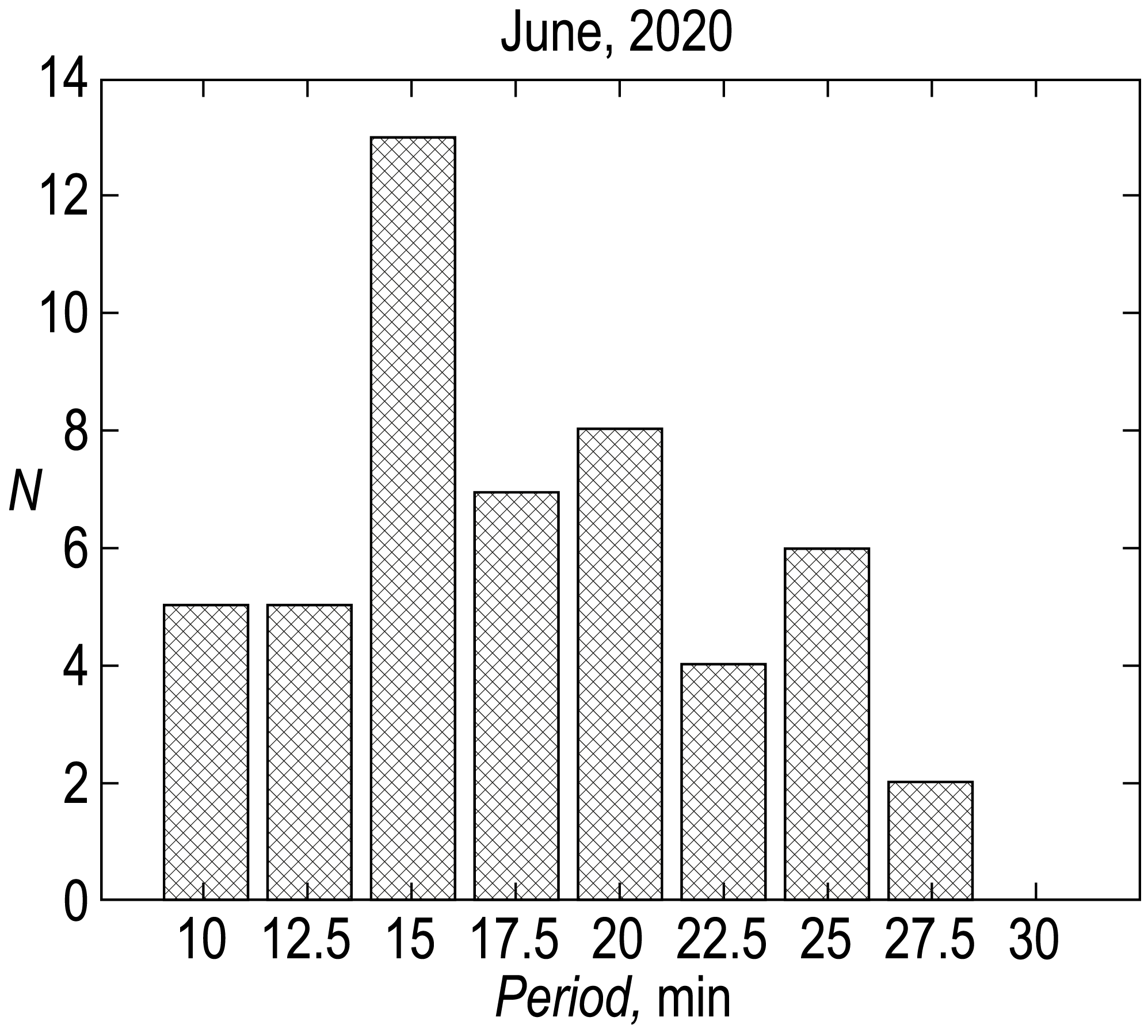}
		\includegraphics[scale=.41]{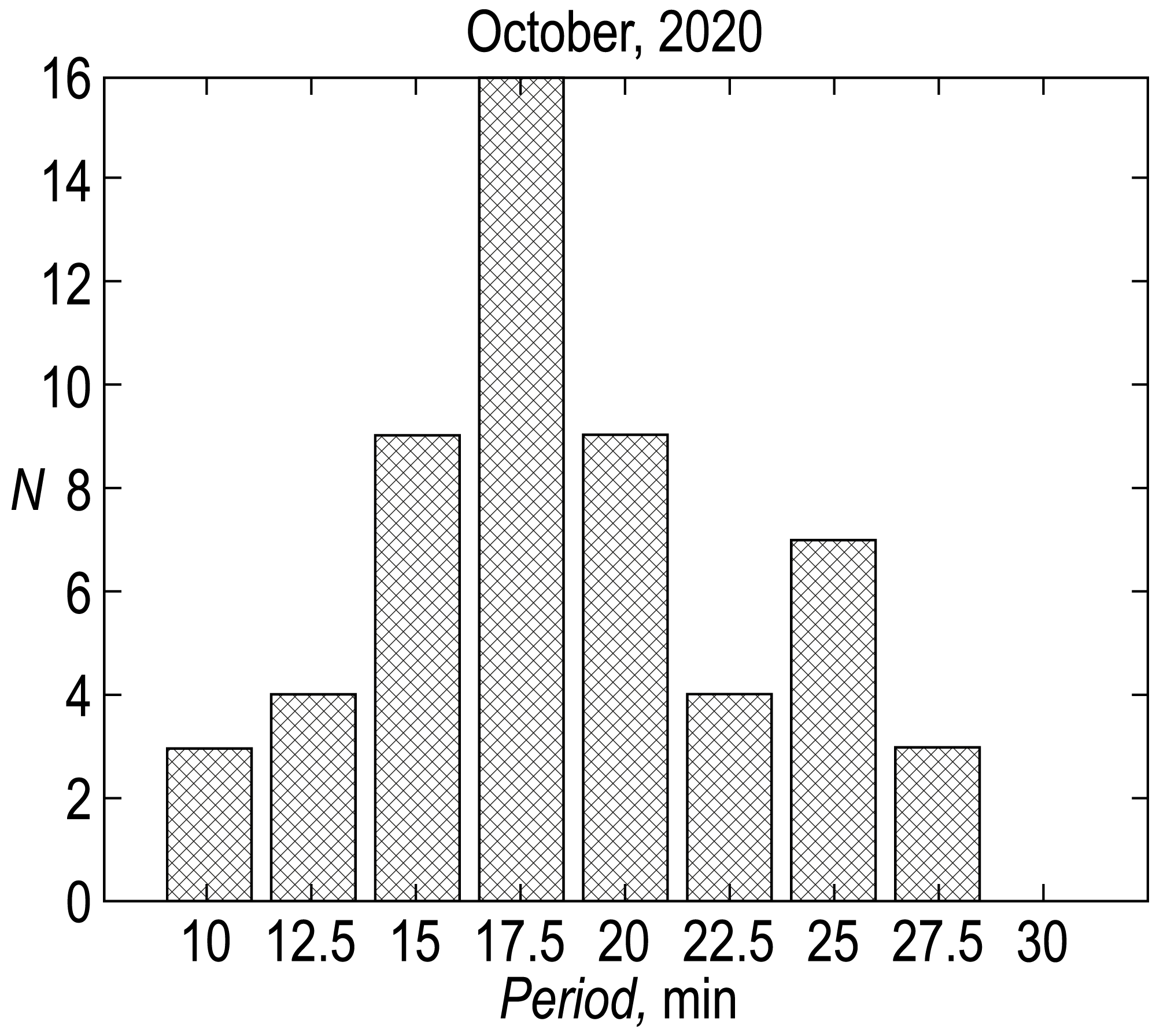}
		\includegraphics[scale=.41]{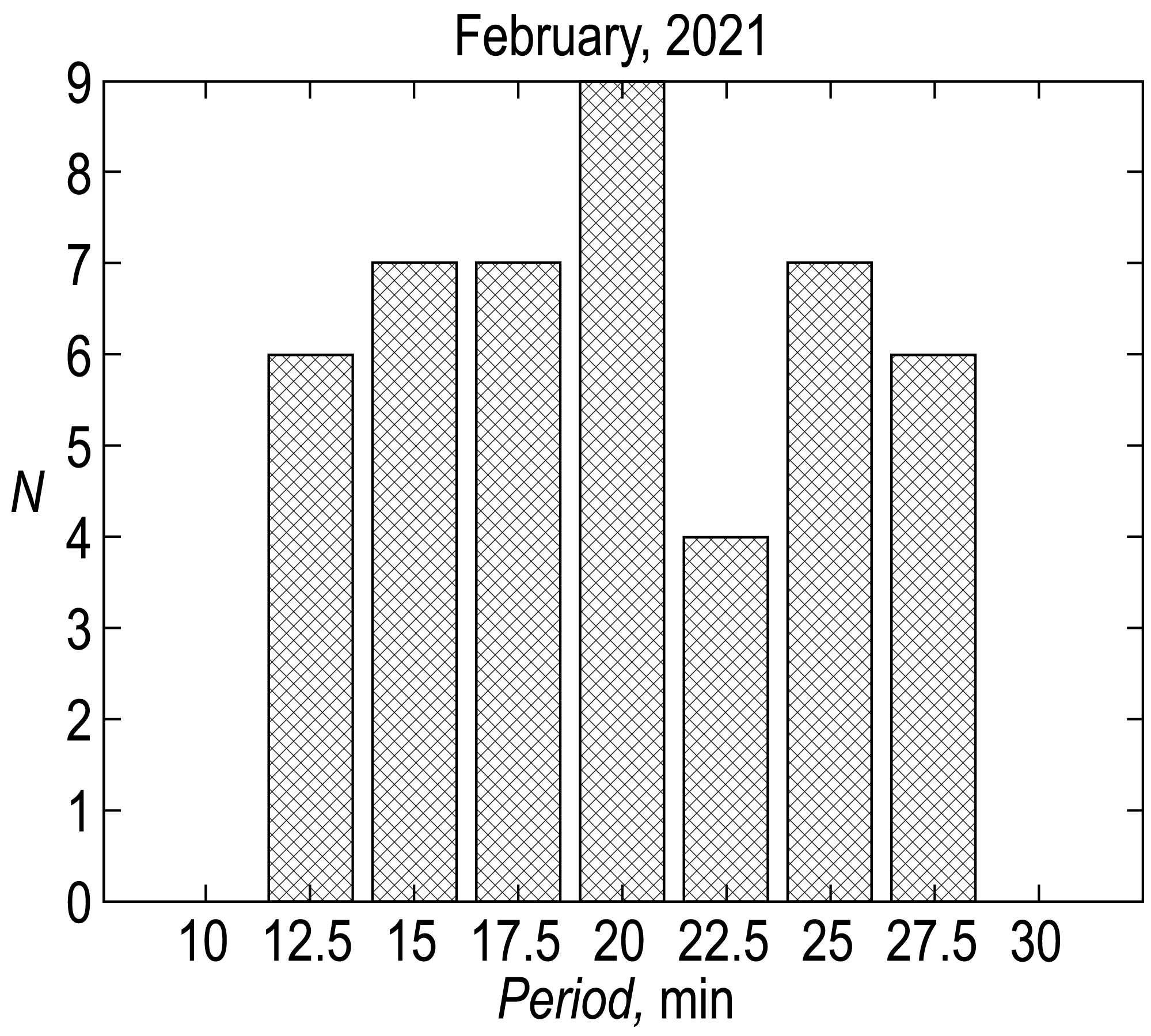}
	\caption{Histograms showing the number of wave disturbances on the terminator with relative amplitudes more than 0.03 depending on the period along the GQD--A118 path in the evening hours. The four panels display the results for the events taking place during April, June, October 2020, and February 2021. The total number of the events analysed in the particular month is the following: 39 (April 2020), 50 (June 2020), 55 (October 2020) and 46 (February 2021).}
	\label{FIG:3}
\end{figure*}

\begin{figure}
	\centering
		\includegraphics[scale=.41]{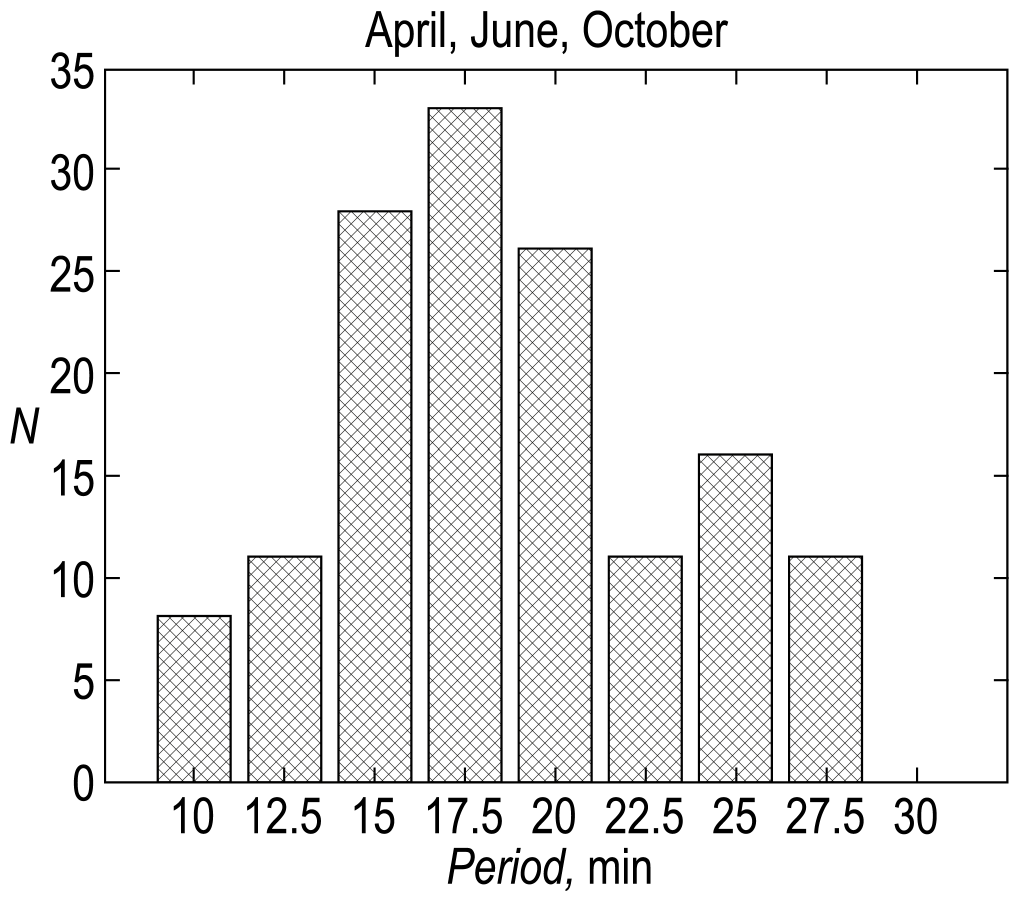}
		\caption{Histogram of the total three-month distribution of periods of waves disturbances with the evening terminator.}
	\label{FIG:4}
\end{figure}

With the help of the method described above, the data of measurements of the radio signals amplitudes along the GQD--A118 path over 4 months (April, June, October 2020, and February 2021) were analysed using Wavelet analysis of evening values. The complex Morlet wavelet was used. Figure \ref{FIG:2} shows the amplitude of the wavelet spectra as a function of time (UT). The results of this analysis are displayed in Fig. \ref{FIG:2} which shows the results corresponding to data covering four separate days related to different seasons. The relative amplitude of fluctuations in different months is typically 3--10\%. It should be noted that large values do not necessarily mean large AGWs amplitudes, but maybe a consequence of changes in atmospheric conditions and the height of radio waves reflection \citep{Fedorenko2021}.
Usually, in the series of measurements of the amplitudes of radio waves, a superposition of oscillations of several time scales is observed, as evidenced by the results of spectral analysis. However, we have noticed that after passing the evening terminator, $\Delta A / \bar{A}$ fluctuations with periods of 15-20 minutes prevail. At the same time, the maximum waves activity develops 1.5--2.5 hours after passing the terminator. In June, waves activity from the terminator is expected to develop later in local time than in other months. Note that along the considered GQD--A118 path, the local solar time is close to UT. 
Therefore, the dependence of radio waves amplitudes on UT is shown in Figs. \ref{FIG:1} and \ref{FIG:2} roughly reflect the dependence on local time.
Figure \ref{FIG:3} shows histograms for the distribution of periods of observed fluctuations with amplitudes limited by the ratio $\Delta A / \bar{A} > 0.03$, plotted for four months from different seasons. For three of the four considered months (except February), the predominance of fluctuations in the range of 15--20 min is noticeable, which indicates the existence of a certain dominant wave mode on the terminator. The diagram shown in Fig. \ref{FIG:4} shows the total distribution of the number of cases of waves fluctuations in three months (April, June and October). In this diagram, the regularities of the distribution by periods are more clearly manifested due to the larger number of events. It is most likely that the wave mode of 15--20 min corresponds to the condition of synchronism with the terminator \citep{Lizunov2009}. In February, the superposition of fluctuations of several scales is more often observed, without a clear predominance of this mode. 
This result may be explained in terms of less solar energy entering the atmosphere or the bad location of the terminator line relative to the radio trace. 
The conditions for observing the main AGW's mode at the terminator should depend on the season and the orientation of the path relative to the terminator line. To study these features, further statistical studies are needed for different seasons with the involvement of measurement data on other radio paths.

\section{Interpretation of observations}\label{s:3}

From general considerations, it is clear that the waves fronts generated by the terminator should move approximately in the direction from west to east due to the rotation of the Earth. These fronts must also be parallel to the terminator line. The horizontal speed of the terminator at the equator at some height, $z$, in the atmosphere, is $V_T=2\pi \left(R_E+z\right)/{T_{rot}}$, where $R_E$ is the Earth's radius and $T_{rot}$ is the Earth's rotation period. This speed is about 460 m s$^{-1}$ near the Earth's surface, increasing to $\sim$ 480 m s$^{-1}$ at an altitude of 300 km. Thus, near the equator, $V_T$ exceeds the speed of sound (see Fig. 6, \cite{Bespalova2016}) near the Earth's surface ($\sim$ 300 m s$^{-1}$) but is less than the speed of sound in the upper atmosphere ($\sim$ 700--900 m s$^{-1}$) at the average solar activity. 
At atmospheric altitudes where the terminator is supersonic (below 150--180 km), both infrasonic and internal gravity waves can be generated \citep{Somsikov1983}. The horizontal speed of the terminator movement decreases in the direction from the equator to the pole approximately according to the law $V_x=V_x (0) \cos \phi$, where $\phi$ is the geographic latitude, and $V_x (0)$ is the speed of the terminator at the equator. For the GQD--A118 mid-latitude path considered in the present study the location of the GQD transmitter is 54.73$^\circ$ N; 2.88$^\circ$ W; (Skelton, Great Britain), while the  A118 receiver is situated at 43.46$^\circ$ N;  1.33$^\circ$ E (Muret, France). 

In the approximation of geometrical optics, the main contribution to the fluctuations of radio waves amplitudes is given by the first ionospheric harmonic, which is reflected from the ionosphere exactly in the middle between the transmitter and the receiver \citep{Yoshida2008}. Along the GQD--A118 path, for this reflection point, the latitude is about 49$^\circ$, that is, the horizontal component of the velocity of the terminator in the middle of the path is $V_{Tx}\approx$ 300 m s$^{-1}$, which is close to the speed of sound. According to the conclusions of the work by \cite{Somsikov1983}, AGWs from the terminator should propagate approximately in the zonal direction with a phase speed that is equal to the speed of the terminator movement, $V_T$, for each harmonic. This is the condition of wave synchronism with a moving source. The condition of wave synchronism with the terminator means that the period and wavelength are related as $T=\lambda_x/V_{Tx}$. Therefore, on the considered radio path, the horizontal phase speed of the waves, $U_x$, which are synchronised with the terminator, is $U_x=\omega /k_x=V_{Tx} \approx$ 300 m s$^{-1}$, where $\omega$ is the frequency, $k_x$ is the horizontal component of the wave vector. 
Therefore, for those AGWs prevailing in the evening hours with periods of 15--20 min, the horizontal wavelengths should be $\lambda_x=U_x T \approx$ 270--360 km. 
Our earlier studies of the satellite data measurements indicate the predominance of AGWs synchronised with it on the terminator  \citep{Bespalova2016}, let us recall some properties of AGWs that follow from this synchronism. The dispersion relation of AGWs can be written as 
\citep{Hines1960,Fedorenko2010}: 
\begin{equation}
k_z^2+{\xi}^2=k_x^2 \left(\frac{\omega_b^2}{\omega^2}-1\right)\left(1-\frac{U_x^2}{c_s^2}\right),
\label{e:1}
\end{equation}
where $k_z$ is the vertical component of the wave vector, $\omega_b$ is the Brunt-V{\"a}is{\"a}l{\"a} frequency, $c_s=\sqrt{\gamma g H}$ is the isothermal sound speed, $H$ is the atmosphere scale height, $g$ is the gravitational acceleration, $\gamma$ is the adiabatic heats ratio, and $\xi^2=g^2\left(1-\gamma/2\right)^2/c_s^4$ is a small value with a wave number dimension ($\sim{\cal O}(H^{-2})$).

At the observation height of these waves, the speed of sound can be calculated as $c_s=\sqrt{\gamma g H} \approx$ 309 m s$^{-1}$, where we used $\gamma=1.4$, $H=7$ km. The Brunt-V{\"a}is{\"a}l{\"a} period, $T_b=2\pi/\omega_b$, at these heights is about 5 minutes, which is 3--4 times less than the prevailing periods of waves observed at the terminator. Therefore, these waves are gravity modes, and not acoustic. Since the speed of the terminator in mid-latitudes is close to the speed of sound, according to Eq.~(\ref{e:1}), the AGWs synchronised with terminator can be either internal waves with a small value of $k_z$ or evanescent waves with $k_z^2<0$ \citep{Cheremnykh2019}. According to the theory, the  horizontal phase velocities of internal gravity waves $U_x<c_s$. Since $\omega_b>\omega$, then at $U_x < c_s$ according to Eq.~(\ref{e:1}) $k_z > 0$ and waves on the terminator propagate freely at some small angle with respect to the horizontal plane. If $U_x > c_s$, then $k_z^2 < 0$ and they propagate horizontally, that is, they are evanescent. However, it is rather difficult to establish exactly whether the observed AGWs belong to freely propagating or evanescent waves due to the closeness of $U_x$ to the speed of sound, which we cannot accurately determine without direct measurements of atmospheric parameters.

\section{Determination of the characteristics of the AGWs on the terminator}\label{s:4}

The propagation of AGWs in the atmosphere is manifested in periodic changes in a number of atmospheric parameters: density, temperature, particle velocity, or pressure. In the case of direct measurements, the amplitudes of AGWs can be understood as amplitudes of fluctuations of any of these quantities, which are related to each other by polarisation ratios \citep{Hines1960}. In this study, we addressed the fluctuations in the amplitudes of VLF radio waves, $\Delta A / \bar{A}$, which indirectly reflect the distribution of AGWs at the heights of the mesosphere where these radio waves are reflected. The periods of the observed $\Delta A / \bar{A}$  fluctuations correspond to the periods of AGWs in the atmosphere, however, information on other spectral and amplitude characteristics of AGWs remains unknown. In the general case, the $\Delta A / \bar{A}$ values are related to the AGWs amplitudes by a complex functional dependence, which is determined to a greater extent by the features of the radio path than by the physical properties of the waves themselves in the atmosphere. Therefore, we can estimate the amplitudes of the AGWs at the heights of radio wave reflection only approximately within the framework of certain assumptions.

\begin{figure}
	\centering
		\includegraphics[scale=.21]{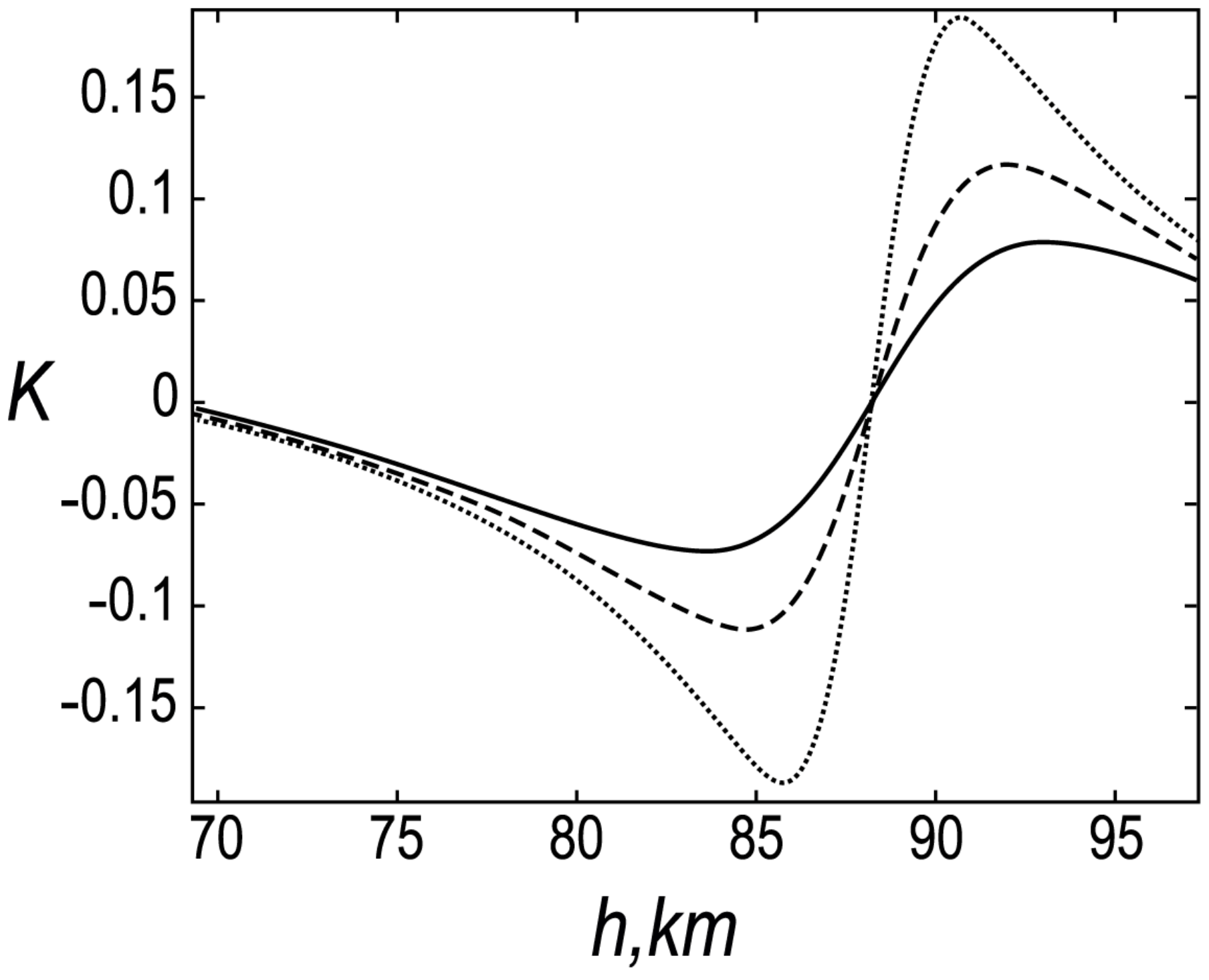}
		\caption{The variation of the coefficient $K$ (km$^{-1}$) on the effective height $h$ of radio waves reflection for three values of the parameter $\beta$ along the GQD--A118 radio path: 2.2 (dashed), 2.5 (hatched), 3 (solid curve).}
	\label{FIG:5}
\end{figure}

To determine the amplitudes of AGWs on the terminator based on the observations of $\Delta A / \bar{A}$ we used the previously developed method \citep{Fedorenko2021}. The authors showed that in the approximation of geometric optics
\begin{equation}\label{e:2}
\frac{\Delta A}{\bar{A}} \approx K \Delta h,
\label{eq:2}
\end{equation}
where $\Delta h$ is the wave fluctuations of the displacement of the effective level of reflection of radio signals, and $K$ is a coefficient that depends on the length of the radio path, the frequency of the signal and the ratio between the amplitudes of near-ground and ionospheric waves. It is clear that the coefficient, $K$, is different for each radio path. The study by \cite{Fedorenko2021} also determined that the expression that relates the relative fluctuations of the concentration of neutral particles, $\Delta N/ N_0$ (in fact, the amplitude of AGWs), with the magnitude of waves fluctuations $\Delta h$ can be written as
\begin{equation}\label{e:3}
\frac{\Delta N}{N_0}=\frac{\Delta h}{H}\left(1+\frac{d H}{d z}\right).
\label{eq:3}
\end{equation}
This relationship confirms that once the value of $K$ for the selected radio path is determined, the amplitude of the AGWs can be calculated in the fluctuations of the concentration of neutral particles $\Delta N / N_0$ from the measurements of the fluctuations of  $\Delta A / \bar{A}$.
By combining Eqs. (\ref{eq:2}) and (\ref{eq:3}) we can determine the amplitude of the AGW as 
\begin{equation}\label{e:4}
\frac{\Delta N}{N_0}=\widetilde{K}\frac{\Delta A}{\bar{A}},
\label{eq:4}
\end{equation}
where
\[
\widetilde{K}=\frac{1}{HK}\left(1+\frac{d H}{d z}\right). \nonumber
\]
For approximate estimates of the amplitude of radio waves at the receiving point along a relatively short path ($<$1500~km), we will consider the interference only two waves: a near-ground wave with an amplitude $A_g$ and the 1st ionospheric wave with an amplitude $A_1$, which is reflected from the ionosphere once before reaching the receiver. As shown by \cite{Fedorenko2021}, in this approximation, the coefficient $K$ can be expressed in terms of the value of the parameter $\beta$ defined as 
\[
\beta=\left(A_1 / A_g\right) + \left(A_g / A_1\right).
\]
It is known that for a radio signal with a frequency of $\sim$20 kHz along 700--1000 km long paths, $A_1\approx A_g$. Along longer paths, the amplitude of the ionospheric waves begins to exceed the amplitude of the near-ground waves due to its lower attenuation \citep{Yoshida2008}. The dependence of $K$ on the effective height of radio waves reflection is plotted in Fig. \ref{FIG:5} for three different $\beta$ values corresponding to realistic values of the $A_1/A_g$ ratio along paths 
 $<$ 1500 km. To clarify the value of $\beta$, we will use the results by \cite{Yoshida2008}, where the values of $A_1$ and $A_g$ were calculated for radio waves of different frequencies and with different track lengths. Considering Fig. 3b in \cite{Yoshida2008}, we obtain that $\beta \approx 2.5$. Therefore, from the dependencies displayed in Fig. \ref{FIG:5}, in the case of the GQD--A118 path, we have to use the variation shown by the hatched curve. It can be seen that for typical daytime radio waves for reflection heights of $\sim$75 km, the value of $K(75$ km) $\approx -0.03$ km$^{-1}$ is several times smaller than the value corresponding to nighttime reflection heights at $\sim$ 90 km, for which $K(90$ km) $\approx 0.1$ km$^{-1}$. If we assume that during the day and at night the AGWs have the same amplitudes of $\Delta N/N_0$, then considering the course of the function $K(h)$, it is clear that the night values of $\Delta A/\bar{A}$, in this case, will be approximately 3 times larger. Accordingly, we expect that in the measurements of amplitudes of radio signals, waves disturbances are more visible at night.

After sunset, there is a sharp transition from day ($\sim$ 75 km) to night heights ($\sim$ 90 km) of the reflection of radio waves. For the GQD--A118 path, the moment of passage of the terminator corresponds to the value of $K(88$ km)$=0$. Note that for different traces the values of height, $h$, at which $K$ passes through zero are different \citep{Fedorenko2021}. To the left of this point of the curve are the morning and daytime $K(h < 88$ km) values, and to the right are the evening and night $K(h > 88$ km) values.

In the evening hours, along the GQD--A118 track, there are fluctuations in the amplitudes of radio waves in the $\Delta A/\bar{A}$ range, which we associate with the propagation of AGWs from the terminator at the heights of the mesosphere. Let us assume that the amplitudes of AGWs, $\Delta N/N_0$, should be approximately the same for several consecutive days since the conditions of illumination by the Sun change by a small amount. Then the observed differences in the evening amplitudes of $\Delta A/\bar{A}$ on a fixed path are caused by a change in the effective height of the reflection height, $h$, which is associated with a change in the background conditions in the mesosphere and ionosphere at an altitude of $\sim$ 90 km. The maximum for four months of observation of the amplitudes of fluctuations in the evening terminator is $\left(\Delta A/\bar{A}\right)_{max}=0.1 \dots 0.12$. These maximum values correspond to $K_{max} \approx 0.12$ km$^{-1}$ and $h \approx 91$ km, which is shown in Fig. \ref{FIG:5}.
Since in the evening and at night the reflection of radio waves occurs near mesopause heights, we will put
$dH/dz=0$ and  the maximum value of the AGWs' amplitude becomes 
\begin{equation}\label{e:5}
\frac{\Delta N}{N_0} \approx \frac{1}{H K_{max}}\left(\frac{\Delta A}{\bar{A}}\right)_{max}.
\end{equation}
Therefore, for the characteristic values of $H=7$ km, $K_{max} \approx 0.12$ km$^{-1}$ and $\left(\Delta A/\bar{A}\right)_{max}=0.1 \dots 0.12$, we will obtain for the amplitude of AGWs in concentration fluctuations $\Delta N/N_0$ = $0.12 \dots 0.14$.

To estimate other characteristics of the AGWs on the terminator, we will use the well-known relationship that follow from the system of hydrodynamic equations \citep{Makhlouf1990}:
\begin{equation}\label{e:6}
\frac{\Delta \rho}{\rho_0}=\left(\frac{\gamma -1}{\gamma}\right)\frac{\Delta z}{H}+\frac{V_x U_x}{c_s^2}.
\label{eq:6}
\end{equation}
Here $\rho_0$ is the undisturbed value of the density, $\Delta z =V_z/i\omega$ is the vertical displacement of the volume element, $V_z$ and $V_x$ are the vertical and horizontal components of the particle velocity. The reflection of radio waves occurs below the turbopause ($\sim 100 km$), where all atmospheric gases are well mixed. Therefore, $\Delta \rho/\rho_0 = \Delta N/N_0$ and Eq. (\ref{eq:6}) is also valid for concentration fluctuations. The first term in the right-hand side of Eq. (\ref{eq:6}) reflects the vertical displacements of the volume element, and the second denotes density fluctuations associated with pressure changes as we can consider that for monochromatic plane waves $\Delta p = \rho_0 V_x U_x$. Since the equilibrium density and pressure in the atmosphere are related by $\rho_0 =p_0\gamma/c_s^2$, the second term can be written as $\Delta p/(\gamma p_0)$. This term actually reflects the contribution of the acoustic volume compression to the resulting density fluctuations. For AGWs with periods $T>T_b$ and horizontal phase velocities small compared to $c_s$, the second term in Eq. (\ref{eq:6}) is usually neglected. However, for AGWs that are in synchronism with the movement of the terminator, the value of the horizontal phase speed ($U_x$) is close to $c_s$, therefore, the acoustic part in Eq. (\ref{eq:6}) cannot be neglected.

Let us estimate the value of $V_z$ using Eq. (\ref{eq:6}) by assuming that $U_x\approx c_s$, $V_x\approx V_z (T/T_b)$, which is carried out for sufficiently large number periods of AGWs compared to $T_b$. After simple transformations it follows from Eq. (\ref{eq:6}) that $\left| \Delta N/N_0\right| \approx \alpha \left|V_z\right|$, where the coefficient $\alpha$ depends on the wave period. Approximate values of the characteristics of AGWs at the evening terminator for two typical periods are given in Table 1. 
\begin{table*}\label{t:1}
\centering
\caption{Characteristics of AGWs on the evening terminator}\label{tbl1}
\begin{tabular}{|l|l|l|l|l|l|}
\hline
AGW period    & Coefficient $\alpha$    &  Fluctuations in particle      & Vertical velocity         & Horizontal velocity      & Vertical volume\\
 min          &  s/m                    &  concentration $\Delta N/N_0$  &  of particles $V_z$, m/s  &  of particles $V_x$, m/s  &  displacement, $\Delta z$, km \\
\hline
15            & $1.6\cdot 10^{-2}$      & $0.12 \dots 0.14$              & $7.5\dots 8.8$            & $22.5\dots 26.4$           & $1.1\dots 1.3$\\
\hline
20            & $2.1\cdot 10^{-2}$      & $0.12 \dots 0.14$              & $5.7 \dots 6.7$           & $22.8 \dots 26.8$          & $1.1 \dots 1.3$\\
\hline
\end{tabular}
\end{table*}

\section{Analysis of the AGWs energetics on the terminator}\label{s:5}

On the GQD--A118 radio path under consideration, the horizontal speed of the terminator is $V_{Tx} \approx 300$ m s$^{-1}$, which is close to the speed of sound. At the same time, for freely propagating AGWs, $U_x<c_s$, while for waves synchronised with the terminator, $V_{Tx}=U_x \approx c_s$. In general it is problematic to determine the type of waves. It should be noted that direct satellite measurements of several atmospheric parameters at the same time make it possible to determine the type of AGWs based on polarisation ratios between fluctuations of various quantities, e.g. density, temperature, and velocity \citep{Klymenko_2021}. In the absence of direct measurements of various parameters to clarify the type of AGWs observed on the evening terminator in the fluctuations of radio wave amplitudes, we will additionally analyse their energetics. It is known that for freely propagating waves, the equality of the average values of kinetic $\bar{E}_K$ and potential energy $\bar{E}_P$ over the period is fulfilled \citep{Fedorenko2010}. If it turns out that $\bar{E}_K \neq \bar{E}_P$, then the observed waves are evanescent. In the case of AGWs we considered the standard dependence \citep{Hines1960}: 
\begin{equation}\label{e:7}
V_x, V_z \sim \exp \left(z/2H\right) exp\left[i\left(\omega t - k_x x -k_z z\right)\right].
\label{eq:7}
\end{equation}
In addition, based on the standard hydrodynamic equations the energy balance equation reads \citep{Fedorenko2010} 
\begin{equation}\label{e:8}
\bar{E}_{Kx}+\bar{E}_{Kz}=\bar{E}_A+\bar{E}_G,
\label{eq:8}
\end{equation}
where the terms denote the average values of the individual components of the AGW\rq{}s energy over the period. Accordingly, $\bar{E}_{Kx}=\rho_0 V_x^2/4$ is the kinetic energy of horizontal movements, $\bar{E}_{Kz}=\rho_0 V_z^2/4$ is the kinetic energy of vertical movements (with 
$\bar{E}_K=\bar{E}_{Kx}+\bar{E}_{Kz}$ the total kinetic energy), 
$\bar{E}_A=\rho_0 V_x^2 \left(\omega/k_x c_s\right)^2/4$ the potential acoustic energy, $\bar{E}_G=\rho_0 V_z^2 \left(\omega_b/\omega\right)^2/4$ the potential thermobaric (or gravity) energy, with $\bar{E}_P = \bar{E}_A+\bar{E}_G$ the total potential energy. For the particular case  of horizontal evanescent wave disturbances \citep[see e.g.][]{Cheremnykh2019}: 
\begin{equation}
V_x,V_z \sim \exp (az) \exp\left[i(\omega t - k_x x)\right].
\label{eq:9}
\end{equation}
In this case, the energy balance equation has the form \citep{Fedorenko2022}: 
\begin{equation}\label{e:10}
\left(a-\frac{N^2}{g}\right) \left(\bar{E}_{Kx}-\bar{E}_A\right) = \left(\frac{g}{c_s^2}-a\right) \left(\bar{E}_G -\bar{E}_{Kz}\right),
\end{equation}
where the value $a$ determines the dependence of the disturbance amplitudes' on height.

It can be seen that at arbitrary values of $a$, the kinetic and potential energies in the case of evanescent AGWs are not equal to each other. The condition of equality of these energies is fulfilled when $a=1/2H$, that is, at the boundary between the region of free propagation and the evanescent region of the AGWs.

Since for waves on the terminator $U_x \approx c_s$, $\omega_b/\omega \gg 1$, then $\bar{E}_{Kx} \approx \bar{E}_A$, $\bar{E}_{Kz} \ll \bar{E}_G$. It can be seen that $\bar{E}_K \neq \bar{E}_P$, that is, the observed waves are evanescent. It follows from Eq. (\ref{e:10}) that with $\bar{E}_{Kx} \approx \bar{E}_A$, the vertical amplitude changes according to the law $a \approx g/c_s^2$. The properties of waves observed at the terminator, $U_x \approx c_s$ and $a \approx g/c_s^2$, correspond to Lamb pseudo-modes \citep{Cheremnykh2019}. Note that this type of waves is generated by the terminator at altitudes $< 100$ km in mid-latitudes, where $V_{Tx} \approx c_s$. Depending on the geographical latitude and height in the atmosphere, the solar terminator can generate different types of waves. Near the equator, at atmospheric heights up to $\sim 100$ km, the horizontal speed of the terminator exceeds the speed of sound ($V_{Tx} > c_s$), so the freely propagating AGWs with $U_x < c_s$ cannot be in synchronism with this source. In the upper atmosphere, we have $c_s \approx 900$ m s$^{-1}$ at the average activity of the Sun, that is, the terminator is subsonic at different latitudes and therefore can generate freely propagating AGWs.

\section{Study of long-term changes in mesosphere parameters}\label{s:6}

It was shown above that the value of $\Delta A/\bar{A}$ is related to the amplitude of AGWs through the complex function $\widetilde{K}$ (see Eq. \ref{eq:4}). This function is determined both by the features of the radio path and by the physical properties of the AGWs, as well as by the state of the atmosphere and ionosphere at the heights of reflection of VLF radio waves. 
When finding the amplitudes of AGWs from measurements of the amplitudes of radio waves an important problem lies precisely in determining the value of $\widetilde{K}$, which depends on a number of parameters. 
\begin{figure*}
	\centering
		\includegraphics[scale=.26]{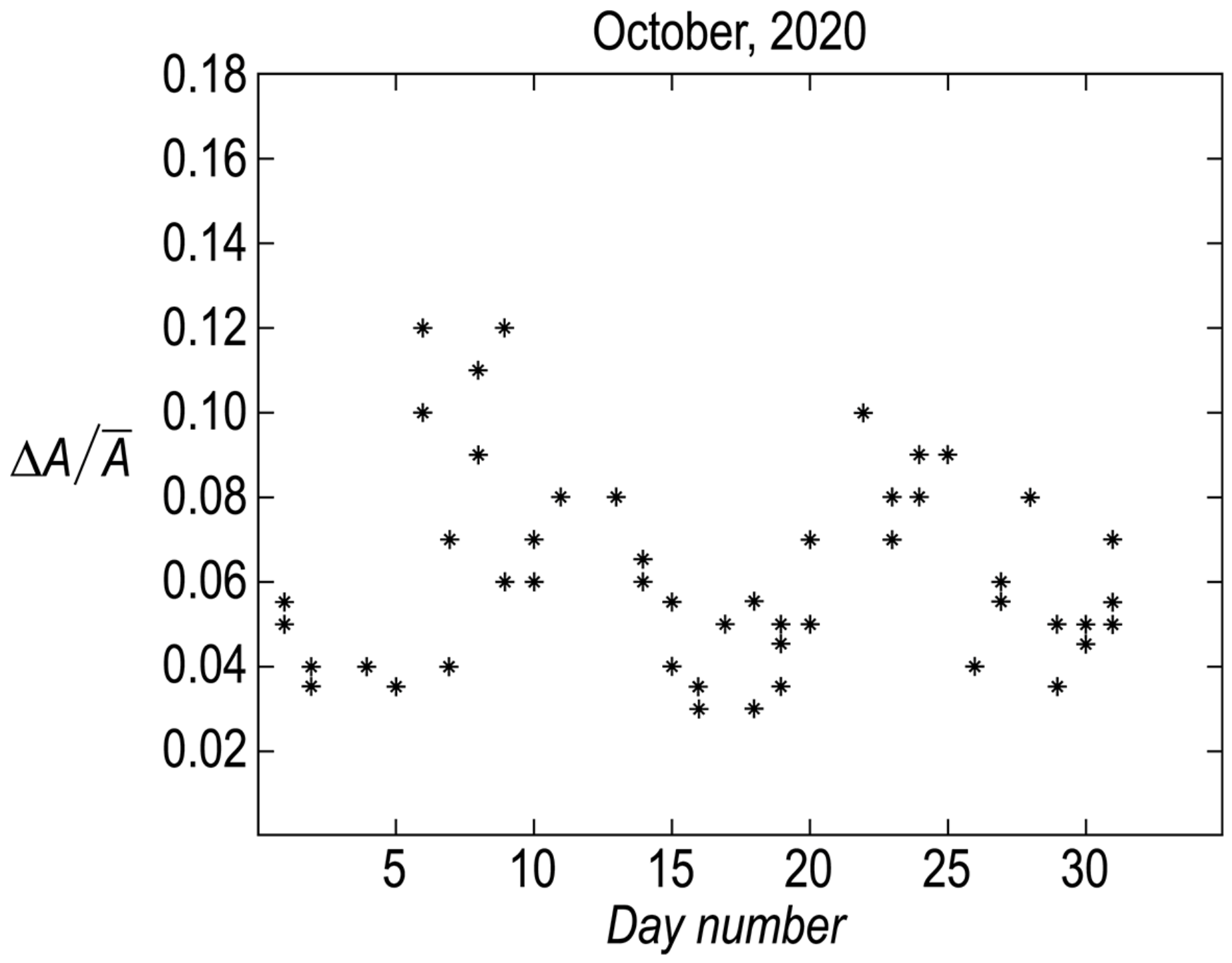}
		\includegraphics[scale=.26]{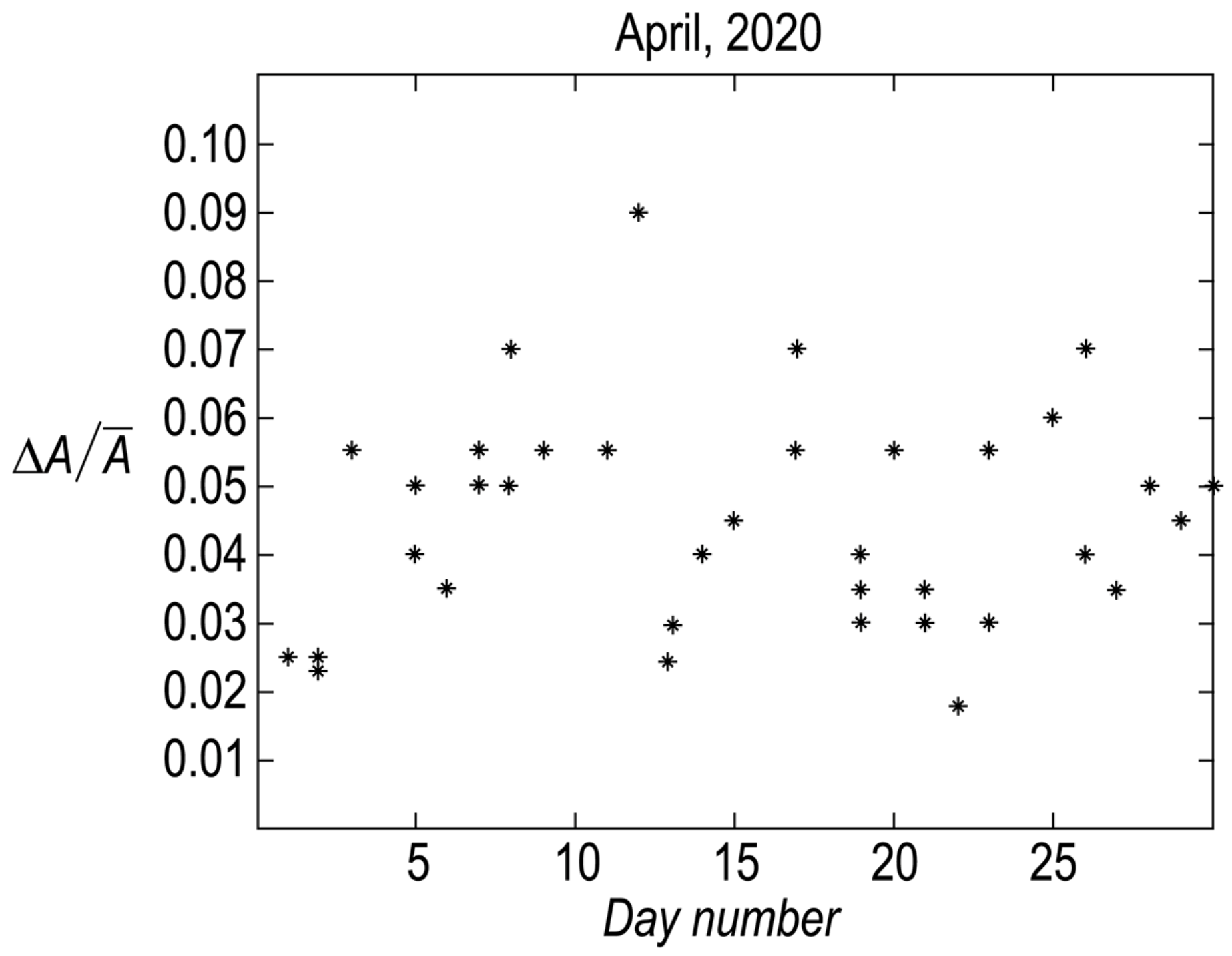}
		\caption{Changes in the amplitudes of AGWs at the evening terminator during October (a) and April (b) 2020.}
	\label{FIG:6}
\end{figure*}
The solar terminator, as a regular source of AGWs, opens up additional opportunities for determining the properties of these waves, as well as for the analysis of long-term changes in atmospheric parameters near the mesopause - at altitudes that are difficult to reach for other observation methods. Since the terminator is a regular source, the properties of the waves generated by it should be similar if the conditions of illumination by the Sun change by a small amount. Therefore, the AGWs on the terminator can be considered as reference waves, the amplitudes of which $\Delta N/N_0$ on a fixed radio path, as well as on different, but geographically close paths, differ by a little.

Our studies revealed that the $\Delta A/\bar{A}$ fluctuations on the terminator along the radio path of GQD--A118 were situated in the interval $\Delta A/\bar{A} = 0.03 \dots 0.12$. Let's suppose that the $\Delta N/N_0$ on the terminator is approximately the same for several days. Then the differences in fluctuations $\Delta A/\bar{A}$ for the amplitudes of radio waves should be related to changes in the state of the atmosphere. Considering the AGWs at the terminator as reference waves, allows us to identify long-term (compared to the periods of the AGWs) trends in the parameters of the atmosphere. These can be seasonal or other changes in the state of the mesosphere, as well as long-period wave fluctuations. On a fixed path, slow changes in the state of the atmosphere at the heights of radio waves reflection will be reflected in the measurement data in the form of slow fluctuations of $\Delta A/\bar{A}$.

An example of such slow changes is shown in Fig. \ref{FIG:6}, which shows the variation of the amplitude associated with $\Delta A/\bar{A}$ at the evening terminator along the path GQD--A118, during October and April 2020. In the data covering October 2020, a quasi-wave structure with a period of $\sim 10-12$ days is clearly visible, which probably represents a planetary waves. In the data for April 2020, there is also a certain trend of $\Delta A/\bar{A}$ amplitudes, but the expressed periodicity is not followed. That is, by measuring the amplitudes of fluctuations of $\Delta A/\bar{A}$ on the terminator, it is possible to study not only AGWs directly caused by this source but also planetary waves manifested in slow trends of $\Delta A/\bar{A}$. For such studies, it is necessary to stitch series of $\Delta A/\bar{A}$ fluctuations on a fixed radio path for several months or even years.

If we consider different, but geographically close radio paths, the changes in the value of $\Delta A/\bar{A}$ on the evening terminator on the same calendar date will differ in amplitude due to the different values of $\widetilde{K}$ on these paths. An example of long-term changes along two closely located radio paths DHO38--A118 (Germany--France) and GQD--A118 (Great Britain--France) is shown in Fig. 5 in the study by \cite{Fedorenko2021}. This figure reveals a slow trend (most likely of seasonal origin) of $\Delta A/\bar{A}$ amplitude fluctuations in the evening hours ($UT = 20\dots24^h$) during 2013–2014. Note that on one of these paths, (DHO38--A118), the amplitudes of $\Delta A/\bar{A}$ are systematically larger, but long-term changes are consistent on both paths. Therefore, when analysing synchronous measurements of $\Delta A/\bar{A}$ values simultaneously on several radio paths, it is possible to specify the value of the coefficient $\widetilde{K}$ for different paths.

\section{Conclusions}\label{s:7}

The wave disturbances from the evening terminator were studied in the range of periods of medium-scale AGWs from ~5 minutes up to 1 hour. Data from measurements of amplitudes of VLF radio waves on the mid-latitude path GQD-A118 (Great Britain-France) were used. A systematic increase in the amplitudes of wave fluctuations was recorded on the path in question for several hours after the passage of the evening terminator. This indicates an increase in waves activity at the heights of the mesosphere, where radio waves of the VLF range are reflected.
Fluctuations in radio signal amplitudes were observed for four months: April, June, October 2020 and February 2021. For different seasons, the existence of predominant wave periods of $\approx15-20$ min at the terminator was found. The obtained results probably indicate the predominant realisation on the solar terminator of wave harmonics corresponding to the condition of synchronism with this moving source.
Amplitudes of AGWs from the terminator at the heights of the mesosphere (relative fluctuations in the concentration of neutral particles, velocity fluctuations, and vertical displacement of the volume) were calculated from the fluctuations of the radio signal amplitudes. The amplitudes of acoustic-gravity waves at the terminator are 12-14\% in relative concentration fluctuations, which correspond to the vertical displacement of the atmospheric gas volume of 1.1–1.3 km.
The energy balance of the AGWs observed at the terminator was analyzed. Based on the energy analysis, it was concluded that in the mid-latitude mesosphere, the solar terminator mainly generates Lamb pseudo-modes with $U_x \approx c_s$ and $a\approx g/c_s^2$.
The possibility of studying long-term changes in the parameters of the mesosphere based on observations of trends in fluctuations of radio waves amplitudes at the terminator is considered.

\section*{Acknowledgments}
The study was supported by the National Research Fund of Ukraine, project 2020.02/0015 Theoretical and Experimental Studies of Global Disturbances of Natural and Man-Made Origin in the Earth–Atmosphere–Ionosphere System. OC, VF, IB and GV are grateful to The Royal Society, International Exchanges Scheme, collaboration with Ukraine (IES$\backslash$R1$\backslash$211177).

\bibliographystyle{jasr-model5-names}
\biboptions{authoryear}
\bibliography{VLF_AGW}

\end{document}